# Disparity of Tropospheric and Surface TemperatureTrends: New Evidence

David H. Douglass[1]*, Benjamin D. Pearson[1], S. Fred Singer[2], Paul C. Knappenberger[3], and Patrick J. Michaels[4]

1. Dept of Physics and Astronomy, University of Rochester, Rochester, NY 14627

2. Science & Environmental Policy Project and University of Virginia, Charlottesville, VA 22903

3. New Hope Environmental Services, Charlottesville, VA  22902

4. Dept of Environmental Sciences, University of Virginia, Charlottesville, VA  22903

*corresponding author. douglass@pas.rochester.edu

**Abstract**

Observations suggest that the earth's surface has been warming relative to the troposphere for the last 25 years; this is not only difficult to explain but also contrary to the results of climate models. We provide new evidence that the disparity is real. Introducing an additional data set, R2 2-meter temperatures, a diagnostic variable related to tropospheric temperature profiles, we find trends derived from it to be in close agreement with satellite measurements of tropospheric temperature. This suggests that the disparity likely is a result of near-surface processes. We find that the disparity does not occur uniformly across the globe, but is primarily confined to tropical regions which are primarily oceanic. Since the ocean measurements are sea surface temperatures, we suggest that the disparity is probably associated with processes at the ocean-atmosphere interface. Our study thus makes unlikely some of the explanations advanced to account for the disparity; it also demonstrates the importance of distinguishing between land, sea and air measurements.

**1. Introduction**

The question of the degree to which Earth's surface temperature is increasing is a climate problem of great interest. The pattern and magnitude of current and /or future warming has both ecological and economic implications. However, the science is not settled on these issues, as many outstanding questions remain. For example, General Circulation Models (GCMs) predict that as a result of enhanced greenhouse gases and atmospheric aerosols, there should be a warming trend that is greater in the low-to-middle troposphere than over the earth's surface [*Chase et al.*, 2004]. However, temperature observations taken during the past 25 years do not verify this GCM result [*Douglass et al.* 2004]..

The globally averaged surface temperature (ST) trend over the last 25 years is 0.171 K/decade [*Jones et al.*, 2001], while the trend in the lower troposphere from observations made by satellites and radiosondes is significantly less, with exact values depending on both the choice of dataset and analysis methodology [e.g., *Christy et al.*, 2003, *Lanzante et al.*, 2003]. This disparity was of sufficient concern for the National Research Council (NRC) to convene a panel of experts that studied the "[a]pparently conflicting surface and upper air temperature trends" and concluded, after considering various possible systematic errors, that "[a] substantial disparity remains"[National Research Council, 2000]. The implication of this conclusion is that the temperature of the surface and the temperature of the air above the surface are changing at different rates due to some unknown mechanism.

A number of studies have suggested explanations for the disparity. *Lindzen and Giannitsis* [2002] have ascribed the disparity to a time delay in the warming of the oceans following the rapid temperature increase in the late 1970s. *Hegerl and Wallace* [2002] have concluded that the disparity is not due to El Nino or cold-ocean-warm-land effects. Other authors [*Santer et al.,* 2000] have suggested that the disparity is not real but due to the disturbing effects of El Niño and volcanic eruptions, a conclusion that has been critiqued by *Michaels and Knappenberger* [2000]. Still others argue that the disparity results from the methodology used to prepare the satellite data [*Fu et al., 2004, Vinnikov and Grody, 2003*]; however, only the results from Christy et al. [*2000*] have been independently confirmed by weather-balloon data [*Christy et al., 2000*, *Christy et al.*, 2003, *Lanzante et al.*, 2003, *Christy and Norris, 2004*].

In this paper, we explore the geographic patterns of the difference between the trends in surface and lower-tropospheric temperatures. We rely not only on surface and satellite temperature measurements for this comparison, but additionally, we employ a set of data, not previously considered, which represents an attempt to reduce tropospheric observations to surface temperature values. Through this methodology, we hope to shed more light on the nature of this disparity.

**2. Data**

We incorporate three temperature datasets into this analysis: observations taken at the earth's surface [*Jones et al.*, 2001], observations of the lower atmosphere made from satellites [*Christy et al.*, 2000], and calculated near-surface temperatures (R2-2m), a diagnostic variable derived from atmospheric temperature observations tied to weather balloons [*Kanamitsu et al.*, 2002]. Each of these datasets contributes unique information to the understanding of the evolution of patterns of temperature at and near the earth's surface.

The "surface" temperature (ST) observations commonly utilized in research and the media are a combination of near-surface air temperatures for land coordinates and below-surface water temperatures for ocean coordinates. The data are monthly anomalies from the 1961-1990 mean temperature within 5º by 5º grid cells.The amount of available data varies with time and grid cell such that some locations have either no data, or spotty data coverage resulting in a total coverage, that is less than global in extent.

The satellite data are observations taken by the microwave sounder units (MSU) [*Christy et al*., 2000]. In this study, we use the MSU data that is best representative of the lower troposphere. These data are monthly anomalies from the 1979-1998 mean values for 2.5º by 2.5º grid cells with complete global data coverage.

Our third dataset is the "2-meter" temperature product (R2-2m) from an update of the original National Centers for Environmental Prediction—National Center for Atmospheric Research (NCAR) reanalysis [*Kanamitsu et al*., 2002; *Kalnay et al*., 1996]. The R2-2m temperature data are modelled primarily from a collection of atmospheric measurements from weather balloons and satellites There is little influence from surface thermometers [*Kistler et al*., 2001; *Kalnay et al*., 2003], although other surface processes, such as snow cover, can contribute. The time-evolution of the R2-2m temperature variable is independent of the MSU-derived lower tropospheric temperatures. The globally complete R2 data begins in 1979 and continues through the near present. However, a change in the snow cover measuring system in late 1998 has resulted in break points in the 2-m temperature series in grid cells with seasonal snow cover (W. Ebisuzaki, personal communication, 2004).

## 3. Methods

Since we wish to examine the disparity in the temperature trends among these three datasets, we limit our analysis to a common observational time series. The starting point in our analysis will be 1979, which is the beginning year in both the R2-2m and MSU data. We truncate the analysis at December 1996 which avoids the snow cover issue in R2-2m. This also avoids the anomalously large 1997 El Nino event in the tropical

Pacific which *Douglass and Clader* [2002] showed can severely affect the trend-line. We will show later in this paper that it is likely that our conclusions would change little had we been able to use data though 2003.

For the period 1979 through 1996, we perform a simple least-squares regression analysis through the monthly temperature anomalies for each grid cell in the R2-2m and MSU datasets (which contain no missing data). For the ST data, however, we must be concerned with missing data. We therefore first aggregate the monthly data into annual temperature anomalies requiring at least 9 months of valid data to produce a valid year, and then perform our trend analysis on those grid cells with at least 16 (out of 18) valid years. We then compare the trends across the three datasets grid cell by grid cell, in latitudinally averaged bands, and in global aggregate. In our comparisons involving aggregated grid cells, we first mask out the trends in the gridcells of the globally complete R2 and MSU data in which there is not valid ST data so that all our comparisons are made to a common geographical area.

## 4. Results
### 4a. Maps of trend-lines for MSU, ST, and R2-2m.

For each cell on the surface of the Earth we show the trend-line for the period 1979-1996 for the MSU, ST, and R2-2m data (See Figures 1A, 1B, and 1C.). One of the most striking observations is that the values are geographically highly non-uniform due in part to the relatively short period examined and the magnitude of natural variations therein.. The greatest positive trends in the Northern Hemisphere (NH) reach values higher than the greatest positive trends in the Southern Hemisphere (SH), and the greatest positive trends are in the mid-latitudes bands. In the NH the highest trend-line values are localized in three areas: Region 1. Netherlands/Germany; Region 2. Manchuria/western Pacific and Region 3. Pacific ocean/western Alaska.

There are no regions of large positive trends in the equatorial band, (nor at the poles for the MSU data) The polar views dramatically show symmetry about the poles. The MSU plots show an unmistakable 3-fold symmetry in the north-polar view and a 4-fold symmetry in south-polar view. While less clear, the same symmetry exists also in the R2-2m and in the ST maps.

Averages computed from these plots are listed in Table 1.

**4b. Latitude dependence of the zonal average**

We compute the zonal averages of the MSU, ST and R2-2m trends for each 5° latitude band and present our results in Figure 2. They all have maximum values close to each other in the band from 40°N to 50°N. However, in the tropics, the MSU and R2-2m trends agree and are both negative, whereas the ST trend values are positive. This difference of ~ 0.2 K/decade at tropical latitudes has been noted before [e.g. *Gaffen et al*., 2000; *Singer* (2001); *Christy et al*., 2001]. It is clear from this graph and the maps that the maximum near 45°N is real and that values are decreasing as one goes towards the pole—in contrast to what some climate models predict. There is also a relative maximum in the SH located at approximately 25°S to 30S°. In addition, we made similar plots of the data with either the land or the oceans masked out. We found that the maximum at about 45N was 50% higher for the ocean only data. This is consistent with the map showing high positive trends over the Pacific extending from Region 2 to Region 3.

### 4c. Northern mid-latitudes (35N to 60N)

Fig. 3A shows latitude band averages (from 35°N to 60N) of the trend values vs. longitude. One sees that the variations in amplitude of all three data sets are of about the same magnitude and phase. We note that the three 'warming' regions defined in Section 4a are readily apparent and that Regions 2 and 3 are connected across the Pacific ocean. From the general agreement in amplitude and phase of these three data sets we infer that the methodologies of all are essentially correct and free from harmful errors.

### 4d. Tropics (20S to 20N)

Fig 3B shows a plot of trend lines vs. longitude, centered at the equator. It is noted that MSU and R2-2m have nearly the same negative means (-0.06 K/decade) while ST is positive (0.09 K/decade) (see Table 1). The difference in mean between ST and MSU/R2-2m is 0.15 and is the disparity noted by the NRC [National Research Council, 2000] and others.

### 4e. Southern mid-latitude band (40S to 20S)

The plots of the three data sets for the SH mid-latitude band are shown in figure 3C. Here the 4-fold symmetry in all three data sets is very noticeable. The averages are: ST: 0.04 K/decade; MSU: 0.02 K/decade; R2-2m: -0.12 K/decade (see Table 1). These differences are smaller than the amplitudes of the 4-fold oscillation so statements about the differences are difficult.

**5. Results and Discussion**

We have studied the temperature trend-lines given by the MSU, R2-2m and ST data for the period 1979 to 1996. There is general agreement among the three (mostly) independent data sets for northern mid-latitudes. It also indicates that the differences we observe in the tropics are real—thus also validating and extending previous results for the tropics that the magnitude of the trend over the oceans is lower than the ST trend [*Gaffen et al., 2000*; *Christy et al.*, 2001].

To assess sensitivity to the length of record, we repeated our analysis of the latitude band average for regions only over the open oceans (regions free of seasonal snow cover issues) for 1979-2002. We found some small changes in the absolute trend values, but the pattern of relative trend differences remained similar thus supporting the robustness of our findings and indicates that our results are not adversely affected by truncating the datasets at 1996.

Our results point to near-surface processes in the tropical regions as a leading cause in the observed disparity between surface and lower tropospheric temperature trends. As most of the tropical region is dominated by ocean areas, it is possible that ocean/atmosphere interactions are a primary driver of the observed trend differences and that sea surface temperatures are not reliable indicators of the overlying near surface air temperatures.

It is interesting to note that the agreement among the three datasets is greatest over the more industrialized northern extratropics, indicating that local processes such as urbanization [*Kalnay and Cai*, 2004] and industrialization [*de Latt and Maurellis*, 2004; *Michaels et al*., 2004] play only a relatively minor role in causing differential vertical temperature trends. This result does not suggest that these processes do not contribute to the observed warming trend, just that they do not contribute greatly to the temperature trend disparity.

**Acknowledgements**

This research was supported in part by the Rochester Area Community Foundation. We thank R. S. Knox for valuable discussions. Additional thanks to V. Patel and Yi-Lun Ding for assisting in some of the computations.

**Table Caption**

Temperature trends from the ST, MSU, and R2-2m data sets in three latitude bands. (1979-1996)

| Trend line (C/decade) | ST | MSU | R2-2m |
|---|---|---|---|
| North(35-60) | 0.224 | 0.244 | 0.228 |
| Tropics(20S-20N) | 0.092 | -0.057 | -0.054 |
| South(20S-40S) | 0.043 | 0.020 | -0.121 |
| global(common area) | 0.106 | 0.027 | 0.014 |
| global (all available data) | 0.106 | -0.005 | 0.015 |

**Figure Captions**

Fig 1. Trend-line maps of ST, MSU, and R2-2m, 1979-1996. North Pole, Full World, and South Pole Projections. For ST, cells with missing data are made dark blue; polar regions for which there are no data are then covered with a colorless circle.

Fig 2. Latitude Plot. MSU, R2-2m, and ST
Zonal averages of trend-lines plotted vs. latitude

Fig 3. Latitude Zonal Average of MSU, R2-2m and ST vs. longitude.

    4A. North. (35N-60N) Note that there are three 'warming' regions: 1. Netherlands/Germany. 2. Manchuria/western Pacific, 3. Pacific ocean/western Alaska. From the maps in Figure 2 one sees that regions 2 and 3 are connected across the Pacific ocean.

    4B. Tropics. (20°S to 20°N).

    4C. South. (40S-20S)

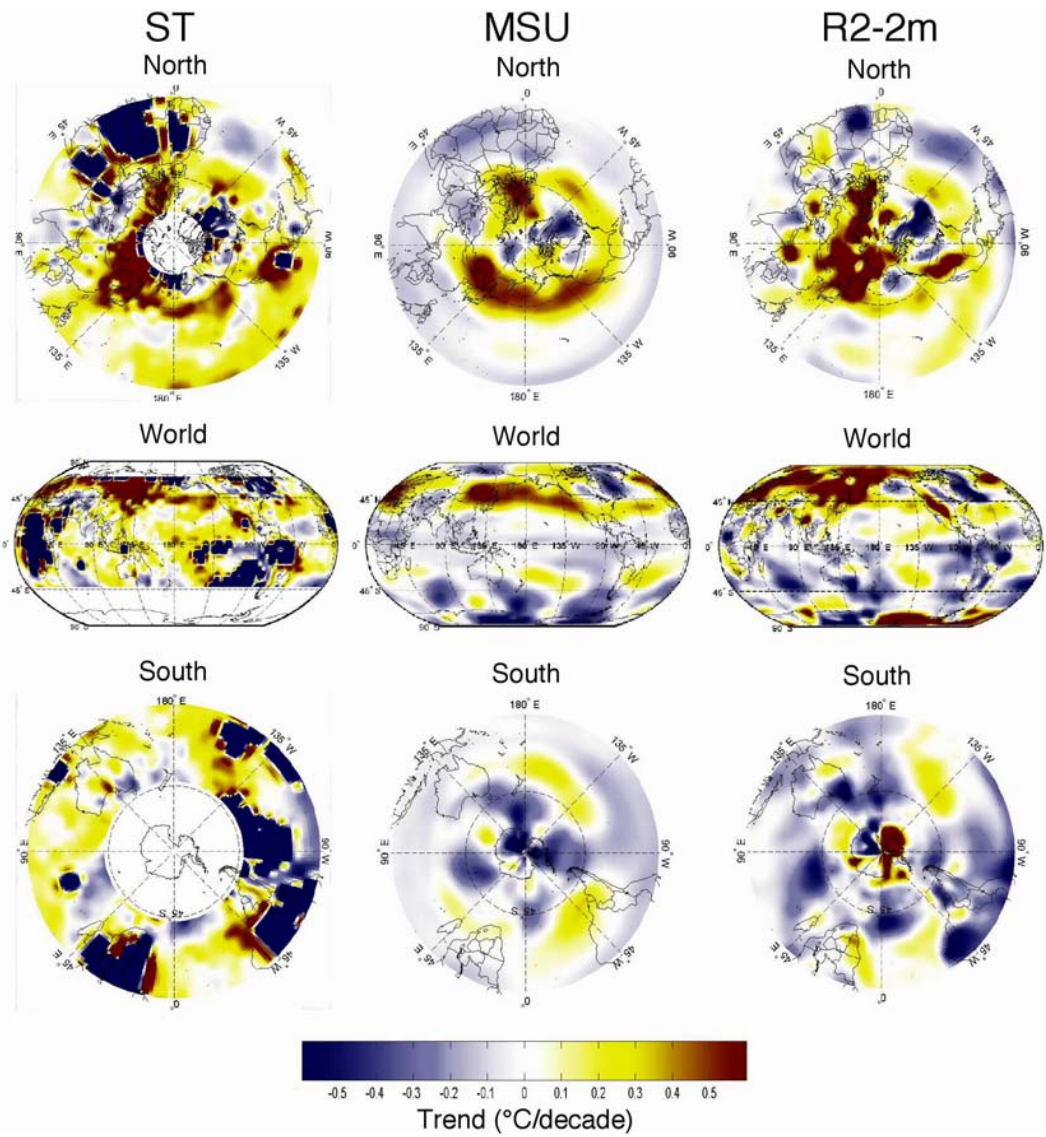

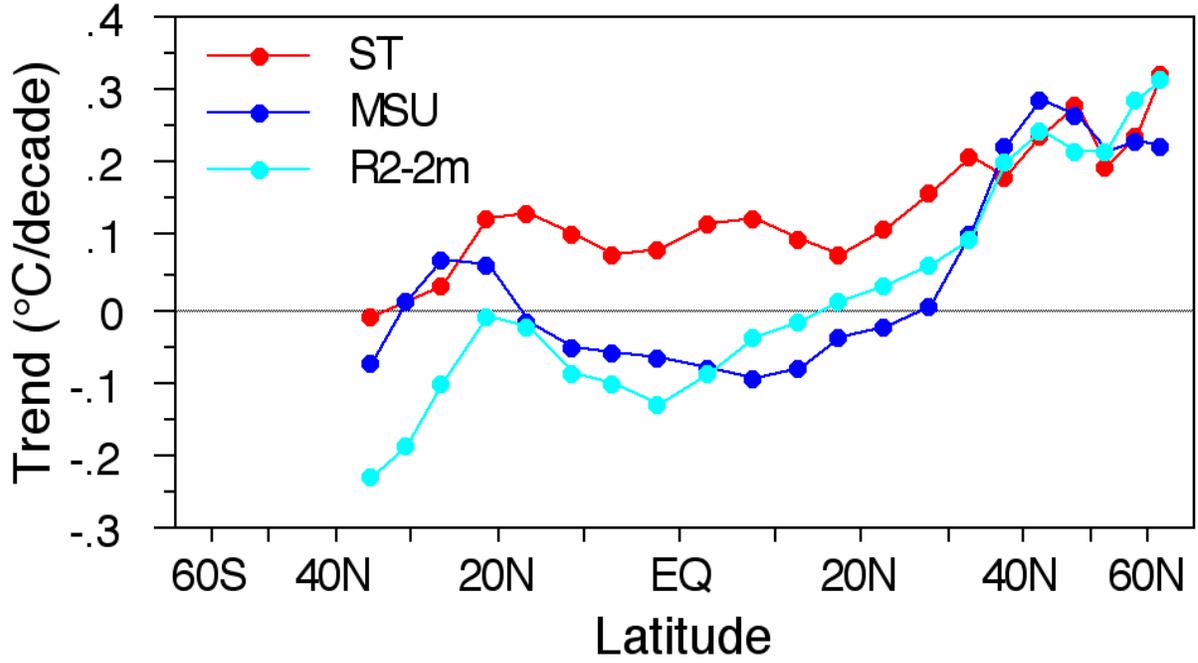

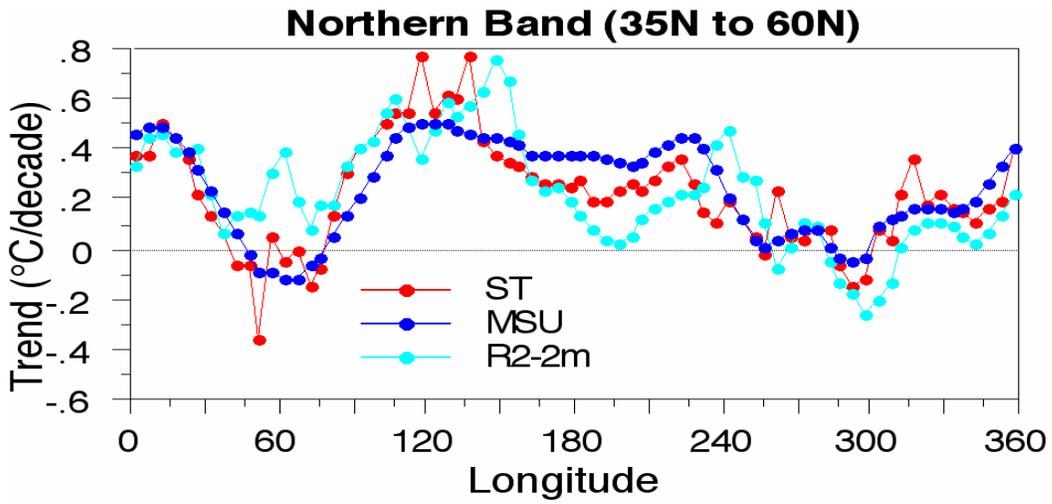

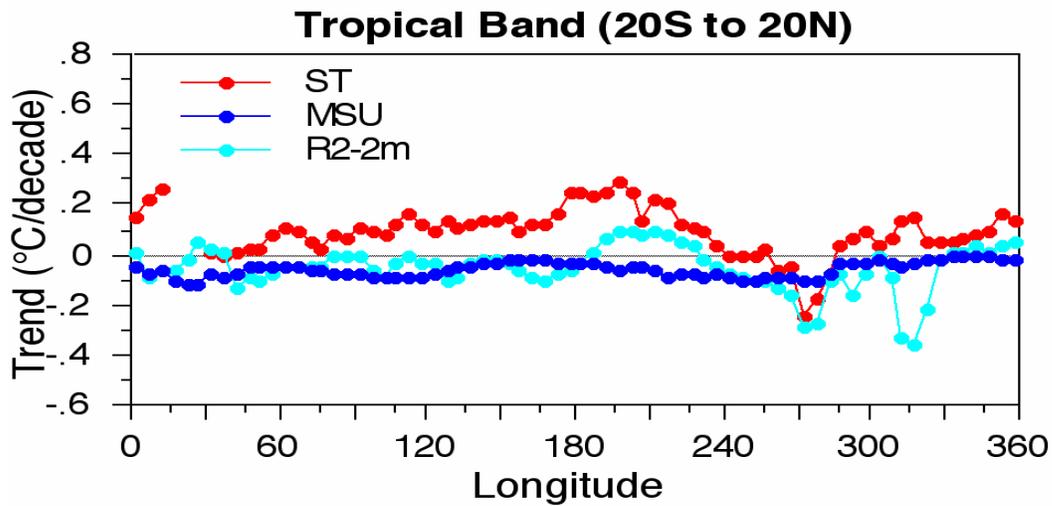

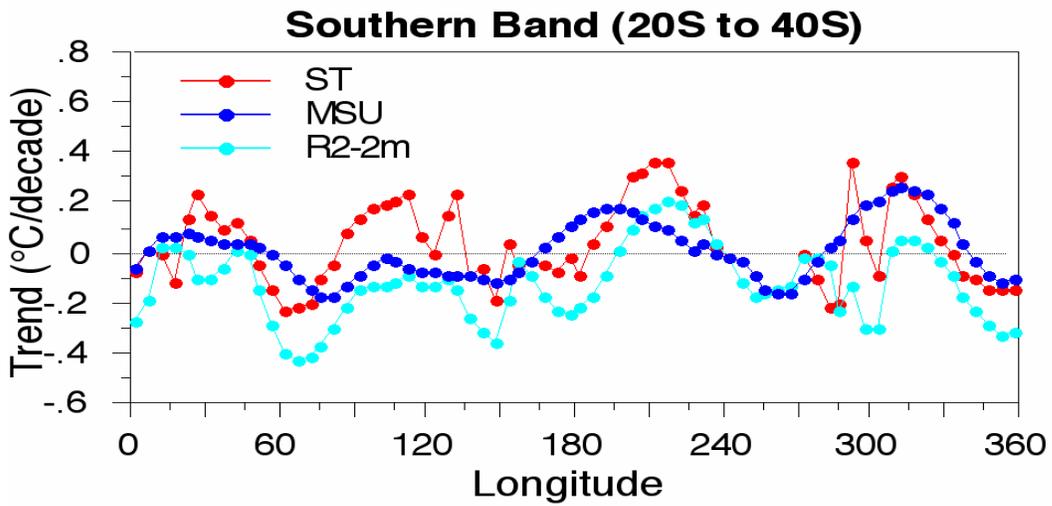